\documentclass[runningheads]{llncs}
\usepackage[T1]{fontenc}
\usepackage{lmodern} 
\usepackage{url}
\usepackage{hyperref}

\usepackage{graphicx}

\usepackage[dvipsnames]{xcolor}
\definecolor{ForestGreen}{rgb}{0.13, 0.55, 0.13}

\usepackage{xcolor}

\begin{document}

\setlength{\parskip}{0pt}

\title{How Do Experts Make Sense of Integrated Process Models?}

\author{
Tianwa Chen\inst{1}\and
Barbara Weber\inst{2}\and
Graeme Shanks\inst{3}\and
Gianluca Demartini\inst{1}\and
Marta Indulska\inst{4}\and
Shazia Sadiq\inst{1}
}
\authorrunning{T. Chen et al.}
\institute{School of Electrical Engineering and Computer Science,
The University of Queensland, Brisbane, Australia\\
\email{tianwa.chen@student.uq.edu.au, shazia@eecs.uq.edu.au, g.demartini@uq.edu.au}
\and
Institute of Computer Science, University of St. Gallen, St. Gallen, Switzerland \\
\email{barbara.weber@unisg.ch}
\and  School of Computing and Information Systems, the University of Melbourne, Melbourne, Australia\\
 \email{gshanks@unimelb.edu.au}
 \and Business School, The University of Queensland, Brisbane, Australia \\
 \email{m.indulska@business.uq.edu.au}
  }

\maketitle              % typeset the header of the contribution

\begin{abstract}
A range of integrated modeling approaches have been developed to enable a holistic representation of business process logic together with all relevant business rules. These approaches address inherent problems with separate documentation of business process models and business rules. In this study, we explore how 
expert process workers make sense of the information provided through such integrated modeling approaches. To do so, we complement  
verbal protocol analysis with eye-tracking metrics to reveal nuanced user behaviours involved in the main phases of sensemaking, namely information foraging and information processing. By studying expert process workers
engaged in tasks
based on integrated modeling of business processes and rules, we provide insights that pave the way for a better understanding of sensemaking practices and improved development of business process and business rule integration approaches. Our research underscores the importance of offering personalized support mechanisms  
that increase the efficacy and efficiency of sensemaking practices for process knowledge workers.

\end{abstract}

\section[Introduction]{Introduction}
\label{Chap:Intro}

The increase in data accessibility and complexity of organisational information systems has given rise to a persistent problem of information silos \cite{vayghan2007internal} wherein knowledge workers have to navigate across different systems and forms of information artefacts to perform their tasks. 
The diversity of information artefacts, ranging from physical to digital, underscores their crucial role in the organization and manipulation of data and knowledge.
For process knowledge workers, including process analysts, process users and process modellers, business process models and business rules are two commonly used artefacts. These two artefacts enable process knowledge workers to represent complex business requirements, as well as implement and improve processes. A notable illustration of the inherent complexity can be observed in the workflows of business analysts and process analysts. For instance, in the scenarios of company mergers and restructurings multiple variants of business processes and business rules need to be consolidated into a single process to eliminate redundancies and create synergies \cite{la2013business}. Even in a business-as-usual environment, if the artefacts are presented separately (i.e., related business rules often may not be part of the business model \cite{zur2008towards}), they may cause a disconnect in shared understanding, and potentially result in conflicts, inefficiencies and even compliance breaches \cite{wang2022business,zhou2023process,zur2008towards}. The challenges facing process knowledge workers involve more than just accessing and understanding information from diverse artefacts. Advanced cognitive and analytical skills are needed to ensure comprehensive understanding, including effectively foraging
and processing various artefacts from business process models and business rule repositories to achieve specific objectives. These foraging and processing processes involve seeking, filtering, reading, and extracting information and iteratively developing a mental model that serves as a foundation for comprehension and performance \cite{pirolli2005sensemaking}.

To overcome these challenges, previous studies have underscored the necessity of comprehensively integrating business rules into business process models \cite{wang2022business}, and various forms of integration have been proposed (e.g. diagrammatic integration, integration through text annotation, and linked rules). 
However, prior research has primarily focused on novice workers using students as proxy \cite{chen2020sensemaking,chen2018business,wang2022business}, which offers limited insights into the sensemaking processes that expert knowledge workers engage in.
Yet, a deeper understanding of how expert knowledge workers forage and process information in integrated process models is key to adequately supporting the development of new process-oriented tools and systems that can more effectively support decision-making processes.

Drawing on the foundational theories of sensemaking and cognition \cite{pirolli2005sensemaking,sweller2011measuring}, our research seeks to delve into sensemaking practices on how knowledge workers forage and process information in the varied forms of integrated representation of business process models and rules. It aims to unearth the underlying factors that drive these various sensemaking behaviours. To this end, we present the outcomes of empirical research conducted  within a controlled laboratory study setting to investigate how process knowledge workers perform tasks based on integrated modelling of business processes and rules (i.e., using text annotation, diagrammatic, and linked rules). Specifically, we investigate experts' sensemaking practices in information foraging and processing phases, and compare the findings with existing literature. By leveraging verbal protocol analysis \cite{gioia2013seeking} with eye-tracking metrics, we reveal empirical insights into knowledge process worker behaviours in information foraging and processing phases.
This exploration paves the way for offering personalized support mechanisms to process knowledge workers through a deeper understanding of sensemaking practices in various settings and improved development of new process-oriented tools and systems.

In the following sections, we first review the research background of sensemaking and cognition as a lens to study process model understanding.  Section 3 introduces our study design and the data analysis methods. Section 4 presents the results and discussion, and finally Section 5 summarizes the contribution of the paper, limitations of the study, and future extensions of this work.

\section[Literature Review]{Literature Review}
\label{Sec:literature}

\subsection{Sensemaking and Cognition}
\label{sec:sensemaking}

Over the decades, sensemaking has been an active area of study in diverse disciplinary backgrounds, from collective organizational contexts (e.g., \cite{weick1995sensemaking,kurtz2003new}) to individual settings (e.g., \cite{naumer2008sense,russell1993cost}). 
More recently, there has been an increased focus on understanding how sensemaking operates in the era of increasingly complex information artefacts (e.g., \cite{Chen2023explor,pirolli2005sensemaking,zhang2020cognitive}). 
Despite the differences between the proposed models from the literature, all attempts describe the iterative process of individual or collective construction of knowledge. A number of models have been proposed to capture sensemaking as multiple loops \cite{pirolli2005sensemaking,zhang2020cognitive}, which consider a fundamental pattern between the interactions of information foraging and processing to schematize the knowledge into a mental model. For example, the Representation Construction Model \cite{pirolli2005sensemaking} has two major loops of sensemaking: (1) the information foraging loop, which includes seeking, filtering, reading, and extracting information processes, and (2) the information processing loop, which includes iterative development of representational schemas to provide a basis for understanding and performance.

The individual settings of sensemaking are more relevant to our work, where the focus is on cognitive mechanisms that underpin individual sensemaking. Cognitive constructs of attention and memory have a natural and strong affinity to the two phases in sensemaking models, and cognitive load theory \cite{sweller2011measuring} provides proven mechanisms through which these constructs can be operationalized \cite{chen2016robust,sweller2011measuring}. For example, attention and search behaviour have been measured through eye-tracking devices, which can capture data on visual scanning (eye movement) and attention (eye fixations) \cite{duchowski2018gaze}. This data, in turn, can be used for various behavioural measurements, such as cognitive load, visual association, visual cognition efficiency, and intensity \cite{bera2019using,rayner1998eye}. 

To the best of our knowledge, existing sensemaking studies are focused on qualitative or perceptive measures with limited use of behavioural and performance measures. Additionally, prior work that used quantitative analysis for studying sensemaking processes was limited to novice workers using university students as proxy
\cite{chen2020sensemaking,chen2018business,wang2022business}. The study of sensemaking practices of novices only provides limited understanding and is not fully reflective of the settings in which these sensemaking processes are undertaken. Hence, we make use of quantitative (eye-tracking devices as observation tool) methods to guide the exploration of qualitative (Cued Retrospective Think-Aloud (CRTA) interviews of expert users \cite{van2005uncovering}) in a controlled 
laboratory study.
%experiment. 
This combination of methods provides novel and objective means to capture and expose sensemaking behaviours and explore the interactive process of how expert knowledge workers forage and process information in various modelling integration approaches.

\subsection{Integrated Modeling of Business Processes and Business Rules} 
\label{sec:ruleIntegration}

Our study considers the specific context of business process and business rule modeling – two complementary approaches for modeling business activities, which have multiple integration methods \cite{knolmayer2000modeling} to improve their individual representational capacity.
The integration methods can be categorized into three approaches with distinct format and construction, namely: text annotation, diagrammatic integration, and link integration \cite{chen2018business}. 
Text annotation and link integration both use a textual expression to describe the business rules and connect them with the corresponding section of the process model. Text annotation integration is a way of representing business rules in business process models by adding textual descriptions of rules – e.g. in BPMN, using the BPMN text annotation construct. With link integration, visual links can explicitly connect corresponding rules with the relevant process section. Diagrammatic integration relies on graphical process model construction, such as sequence flows and gateways, to represent business rules in the process model. Each of these methods has strengths and weaknesses, and thus a potential impact on a knowledge worker’s understanding of a process \cite{wang2022business}. 
Despite the use of rigorous quantitative analysis in related works \cite{chen2020sensemaking,chen2018business,wang2022business}, we found that quantitative analysis alone was not sufficient to fully capture the nuanced behaviors involved in sensemaking. This observation underscores the necessity for rich qualitative insights to thoroughly understand sensemaking behaviors.

\subsection{Process Model Understanding}
Prior research has focused on a variety of factors that affect the understanding of a process, including process model factors \cite{figl2015influence} and human factors \cite{mendling2012factors}.
Process model factors relate to the metrics of the process models, such as modularization \cite{reijers2011human}, block structuredness \cite{zugal2012assessing}, and complexity. Studying the impacts of these involves investigation of the number of arcs and nodes \cite{mendling2012factors}, number of gateways \cite{reijers2011human}, number of events \cite{rolon2008evaluation}, number of loops \cite{figl2015influence}, and number of concurrencies \cite{mendling2008influence}, length of the longest path \cite{mendling2008influence}, depth of nesting \cite{gruhn2006adopting}, and gateway heterogeneity \cite{mendling2008influence}. Human factors  
relate to the factors of process model users, such as individual's domain knowledge \cite{turetken2016effect}, modeling knowledge \cite{figl2015influence} and modeling experience \cite{mendling2012factors}.

To evaluate cognitive engagement and improve comprehension of process models, the think-aloud approach has served as a pivotal means to understand user interactions and cognitive processes during task completion \cite{bera2011guidelines,haisjackl2016understanding,haisjackl2018humans,zugal2015investigating}.
The approach has several methods, including Concurrent Think-Aloud (CTA), Retrospective Think-Aloud (RTA), and Cued Retrospective Think-Aloud (CRTA) \cite{van2005uncovering}.
The CRTA approach integrates elements of RTA and CTA, and mitigates the memory-related limitation in RTA and potential disturbance on task completion of CTA. By ensuring participants' natural interaction patterns are preserved during completing the tasks as well as providing them with concrete and task-specific stimuli cues (e.g. screen recording of completing the task), CRTA can facilitate participants to recall their thought process more accurately. 
In addition, cognitive load and visual cognition have been used as measures of process model understanding \cite{mendling2012factors}, with the use of eye tracking technology to capture eye movement and gaze patterns \cite{bera2019using,petrusel2013eye,petrusel2016task,schreiber2024cognitive}. For example, researchers used eye tracking to 
investigate the visual cues of colouring and layout with performance in process model understanding \cite{petrusel2016task}, and the impact of the task type (local or global) on process model comprehension during information search and inference phases \cite{schreiber2024cognitive}, as well as with the use of RTA to explore reading patterns and the strategies in DCR-HR \cite{abbad2021exploring}.

Upon reviewing the literature, sensemaking emerges as a promising new perspective that has yet to be fully explored in the context of process model understanding. In light of these considerations, our study 
leveraged eye-tracking data as an observational tool and a cue during interviews, guiding participants to reflect on their recorded gaze behaviors, thus enhancing the recall and depth of their thought processes.  This integration facilitates and advances a deeper qualitative exploration into the complex sensemaking behaviors of participants, offering enriched insights into their cognitive engagement with integrated process models, as ``the best cues will likely come from the participants themselves'' \cite{boren2000thinking}.

\section[Study Design]{Study Design}
\label{Sec:studydesign}

In this study, we used a laboratory study method \cite{adams1957laboratory} and a between-subject design with purpose-built platforms.
To capture the insights of sensemaking behaviors, we first collected eye-tracking data while participants performed the laboratory study. We did this to develop a ``cue'' to be used in the main method reported on in this paper --- the cued retrospective think-aloud method (see Section~\ref{sec:sensemaking}), which  involved using a semi-structured protocol for each expert.
Cued with eye gaze movement recordings,  CRTA enabled the participants to verbalize and explain their sensemaking practices and strategies during information foraging and processing. This approach facilitated an in-depth qualitative analysis of the participants' information needs and intentions, as well as the challenges and difficulties they encountered, providing valuable insights into their cognitive processes.

\subsection{Participants}	

Our study is specifically focused on experts. All participants in our study were academic researchers with prior experience as practitioners in business process management. They were from the information systems and computer science disciplines in two universities and a research institute. They were required to have both prior work experience as practitioners and research experience using BPMN of over four years. 
In line with acceptable numbers of participants in 
qualitative studies
\cite{haisjackl2018humans,marshall2013does,nielsen1994estimating}, 15 experts participated in this study, with 5 participants engaging in each integration approach.

\subsection{Laboratory Study Materials and Procedure}

While the main focus of this paper is the rich qualitative data offered by CRTA, the laboratory study data also consists of a pre-study questionnaire, eye tracking data, task performance and log data, and a post-study questionnaire using the NASA-TLX \cite{hart2006nasa} to collect perceived task load. 
Our business process modeling language of choice was BPMN 2.0, due to its wide adoption and its standing as an international standard. The scenarios of the model and rules originated from a car insurance diagram included in OMG’s BPMN 2.0 documentation\footnote{ Model originated from a travel booking diagram in OMG's BPMN 2.0 examples can be viewed in \url{http://www.omg.org/cgi-bin/doc?dtc/10-06-02}}. For expert groups, we used the three integration approaches (one for each treatment group). We ensured, through multiple revisions, that we created informationally equivalent models for all three integration approaches. Due to space limitations, the models cannot be included in the paper, but the complete laboratory study instruments are available for download\footnote
{Please view laboratory study instruments in  \url{https://bit.ly/CAiSE2025}}. 
We ensured all confounding factors were constant. In particular, the model was adjusted to ensure consistency of format for each of the integration approaches (i.e., using text annotation,
diagrammatic, and linked rules). 
In total, there are three questions in the laboratory study. The questions differed in terms of the modeling constructs a participant will have to review to answer them. The model constructs for Q1 and Q2 include sequence and AND gateways, while Q3 include sequence, AND gateways, and XOR gateways. This diversity allowed us to gain further insights into the relationship between integration approaches and task complexity (reflected by the coverage of the model required to answer a particular question). 
Moreover, questions differed in terms of their span and 
each question is related to different process areas and business rules (a participant may have to navigate only a specific section of the process model to answer the question for Q1 and Q2 (local question), or the whole process for Q3 (global question)) (see Section 3.3 on details of process areas related to each question answer).

The complete procedure consists of six steps, described as follows.
(Step 1) We first used a screening form to recruit potential participants. Then we provided a pre-study questionnaire to eligible participants. To ensure group balance, we used a pre-study questionnaire to capture participants' prior knowledge and basic demographics, which we used to distribute participants across groups to avoid accidental homogeneity. 
(Step 2) We set up the laboratory study environment with eye tracking device. For each participant, we provided training on the instructions of the laboratory study, eye tracker device and calibration. 
(Step 3) After calibration, participants were first provided with a BPMN tutorial and were then offered a model using one of the three rule integration approaches. We encouraged each participant to ask questions during the tutorial session, to ensure their readiness for the laboratory study. 
(Step 4) In the laboratory study, all participants had to answer 3 questions.  
We did not set a limit on the laboratory study duration nor a word count limit on participants’ answers. We recorded the gaze movements while experts were working on the tasks.
(Step 5) Upon task completion, participants were provided with a post-study questionnaire using the NASA-TLX 
\cite{hart2006nasa} to collect perceived task load. 
(Step 6) For each participant, we replayed the recording of the task completion process with their gaze movements, and we asked the following semi-structured questions to understand their behaviours in line with our objectives of the study. (1) Would you summarize the process for answering the questions? And explain why you worked in this way. (2) Please watch the replay video, can you identify the point in the video when you knew the answer (indicate the point in time when identifying the answer)? Explain what made you realize the answer, giving specific details (e.g., which activity/rules make you feel you know the answer?) and how confident are you in your answer to this question? (scale 1-5) (3) Regarding the difficulty level of these three questions, did you find any questions that are easier or harder? (4) Overall, can you comment on difficulties or challenges you experience with regard to understanding the model and rules when answering the questions?

\subsection{Setting} 

The eye tracking data was collected through a Tobii Pro TX300 eye tracker\footnote{For more specifications of the eye tracker, please visit
\url{https://www.tobii.com/products} }, which captures data on fixations, gazes, and saccades with timestamps. The laboratory study was conducted in full-screen mode and complete models were displayed without the function of zooming in or scrolling\footnote{Laboratory study design can be viewed and downloaded from \url{https://bit.ly/CAiSE2025}}. The visibility of the text and diagrams was examined carefully, with all text and diagrams being clear from a distance of 1.2 meters. All laboratory studies were conducted in the same lab with the same eye tracker.

\subsection{Analysis Approach}
\label{Sec:analysis}

To uncover the significant differences in knowledge workers' behaviour between the three representations, we conducted the analyses on verbal protocol collected in cued-retrospective think-aloud interviews and complemented them with insights derived from the eye-tracking data. In line with sensemaking foundations (please see Section~\ref{sec:sensemaking}), we segment the laboratory study into two phases, namely the information foraging phase and the task-specific information processing and answering phase.
The information foraging phase for a particular task commences when the participant first fixates on the laboratory study screen, and the information processing phase commences when the participant starts to type the answer in the question area for the first time.

\textbf{Verbal Protocol Analysis.} To analyse the user insights collected from the cued-retrospective think-aloud interviews, we used NVIVO 12\footnote{For the use of Nvivo 12, please view \url{https://lumivero.com/products/nvivo/}} for verbal protocol analysis and we followed the procedure outlined by Gioia et al \cite{gioia2013seeking}, to show the ``dynamic relationships among the emergent concepts that describe or explain the phenomenon of interest and one that makes clear all relevant data-to-theory connections'' \cite{gioia2013seeking}. The verbal protocols were transcribed and provided to two independent coders for analysis to reduce bias in our analysis. 
We started the analysis following an inductive approach and then transitioned into a more abductive approach, to ensure ``data and existing theory are now considered in tandem'' \cite{gioia2013seeking}. The analysis process included four steps. In the first step, we aimed to ``let the data do the talking'', so we inductively analysed the verbal protocols to identify key processes and strategies on how participants foraged and processed information to solve each task. Then we conducted open coding \cite{strauss1998basics} based on the key themes and foci that emerged from the first step, and then the authors discussed the codes with the independent coder iteratively until an agreement on the first-order coding was reached. In the third step, we aimed to answer ``whether the emerging themes suggest concepts that might help us describe and explain the phenomena we are observing'' \cite{gioia2013seeking}. Using existing theories of sensemaking and cognition combined as a theoretical frame of reference, we reflected on the way in which the second-order concepts represented or related to knowledge workers' sensemaking behaviours in information foraging and information processing. We reviewed the existing literature and theories to analyse and develop concepts that explain the data and we did not allow prior theoretical concepts and assumptions to restrict our interpretations.

\textbf{Eye Tracking Analysis.} In this study, the eye tracking data served as a preliminary observation tool that guided deeper qualitative exploration in CRTA interviews into the sensemaking behaviors of participants, where participants could reflect on their recorded gaze behaviors, enhancing the richness of the qualitative data by connecting it to observable actions. 
To reveal the insights on nuanced differences in how experts allocated attention across three types of integration approaches during tasks of varying complexity, we conducted complementary analyses of insights from eye tracking metrics.

\begin{figure}
    \centering
\includegraphics[width=0.9\linewidth]{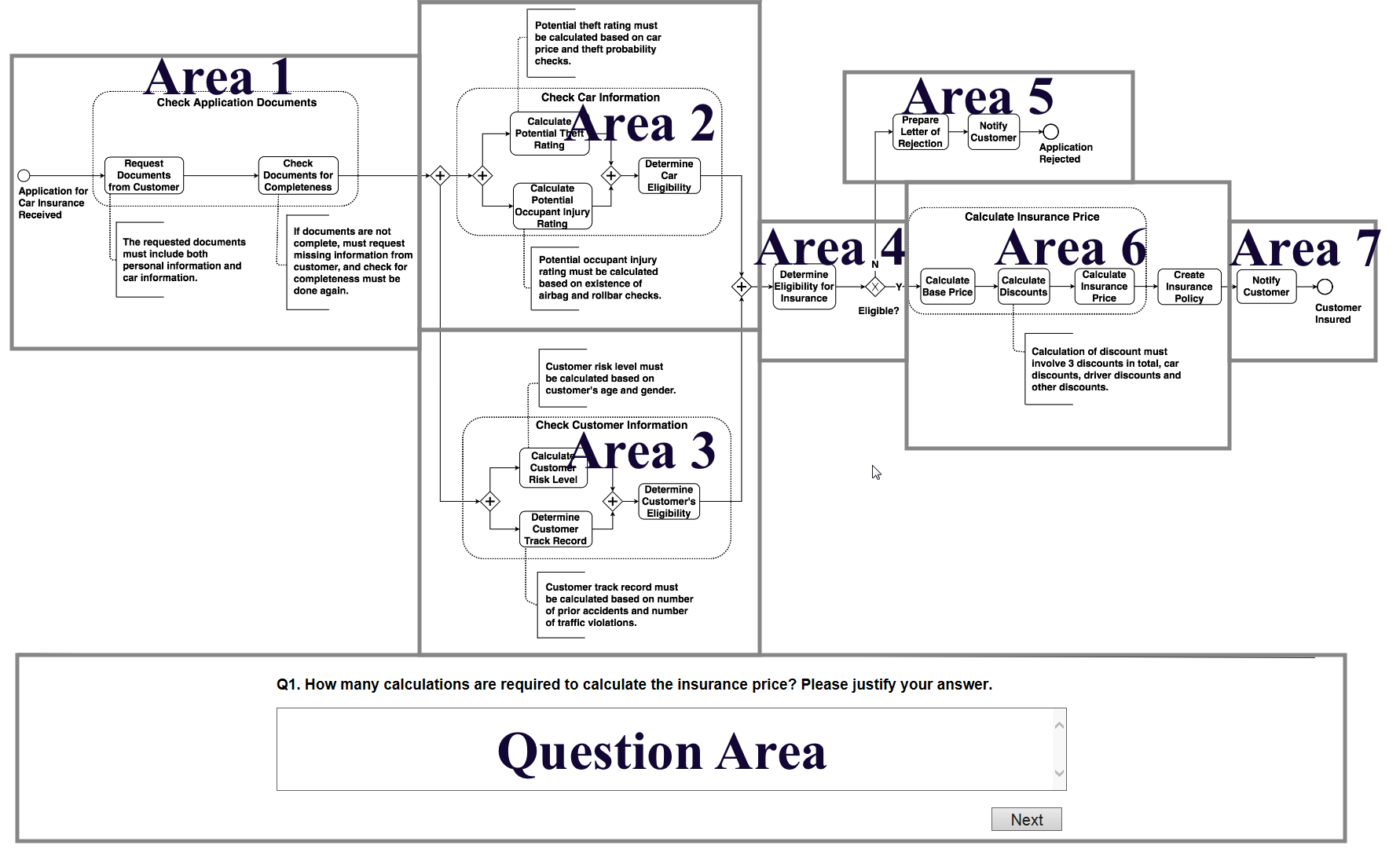}
    \caption{Visual design in laboratory study – text annotation integration approach}
    \label{fig:Experiment platform}
  
\end{figure}

We analyzed the efficiency of locating the answer during the information foraging phase on relevant areas (a measure of visual association of Areas of Interest (AOIs) \cite{bera2019using}). The timestamp of locating the answers is indicated by each expert participant during the cued-retrospective think-aloud session.
The duration of locating the answer begins when the participant starts each question until the timestamp when they indicate they found the relevant information on the model area.
The AOIs were used for analysis and were invisible to participants. 
As shown in Fig.~\ref{fig:Experiment platform}, for models featuring text annotation and diagrammatic integration, the screen was divided into 8 areas: seven different process model areas and a question area (which showed one question at a time). For models featuring link integration, there was an additional ninth area for rules, which displayed the corresponding business rules when participants clicked on each ``R'' icon in the model. Each question answer is related to different process areas. For local questions Q1 and Q2, the answer is related to area 6 and area 2, respectively. For Q3 (global question), the answer is related to areas 1, 5 and 7.
   
\section[Results]{Results and Discussion}
\label{Sec:results}

\begin{figure}
    \centering
    \includegraphics[width=0.75\linewidth]{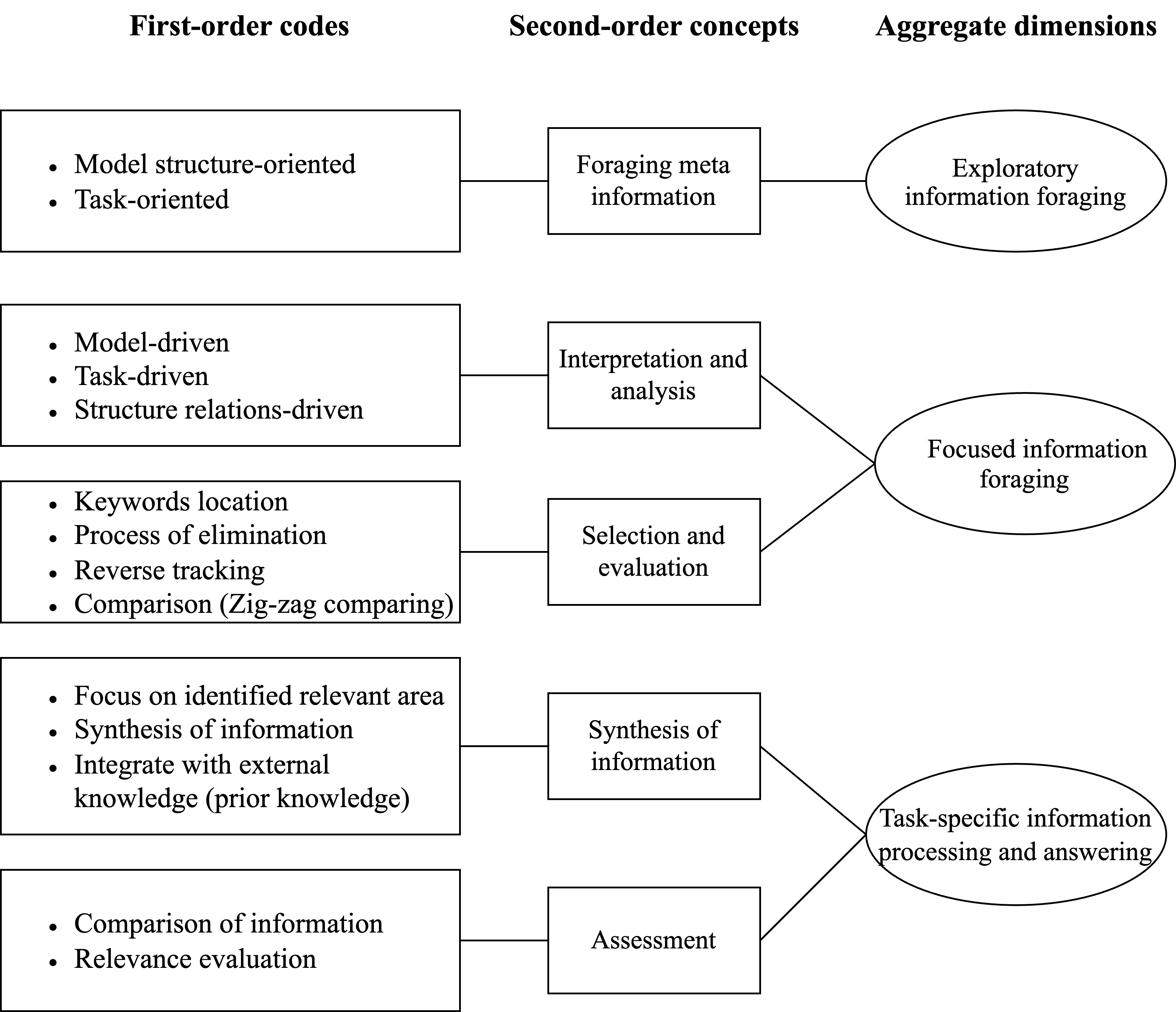}
    \caption{Data structure of expert process knowledge workers' sensemaking practices}
    \label{fig:dataStructure}
\end{figure}

As discussed in Section~\ref{Sec:analysis}, we followed the analysis procedure outlined by Gioia et al \cite{gioia2013seeking}. Our findings shown in the data structure (please view Fig.~\ref{fig:dataStructure}) revealed that expert knowledge workers have distinct sensemaking loops and strategies during exploratory information foraging, focused information foraging and tasks-specific information processing and answering processes. To suit different information needs, they optimized and switched strategies during these processes. We also found these sensemaking practices were not mutually exclusive, and usually were interrelated and embedded with each other and could be used concurrently based on the different information needs. In addition, all participants expressed that using their gaze recordings was an effective approach for them to remind and explain their attention. 

\textbf{Exploratory information foraging. } All the experts (15/15) indicated that at the beginning, they would quickly start scanning either from the business process model (structure-oriented) or the task (task-oriented) to understand the meta-information about the context and structure of the business process model and question. E.g., \textit{``I think, before I even like really worked on the first question. I've had the whole process to get an overview. And then I started going module by module (each activity group).''} (P15). 
Participants explained that they wanted to understand the big picture from the meta-information before zooming into details. 
This insight is also supported by all experts presenting similar attention on the model area during the first information foraging process in Q1 irrespective of the integration group\footnote{The normalized fixations duration in Q1:
Text annotation (M = 0.195, SD = 0.125), Diagrammatic (M = 0.284, SD = 0.087) and Linked (M = 0.280, SD = 0.137). }.

\textbf{Focused information foraging.}  All experts (15/15) indicated that after they understood the meta information they would focus on the details, and we observed that they used various strategies during interpretation, analysis, selection, and evaluation to forage the information on the process model and business rules. 

\textbf{\textit{Interpretation and analysis.}} 
The findings revealed several main strategies for gathering and interpreting information, encompassing task-driven, model-driven, and structure-relations driven approaches during the process of foraging information. 
In the first task (Q1, local question), the majority of the participants (11/15) indicated that they first read through the task and abstracted the key information from the question (\textbf{task-driven strategy}). E.g., \textit{``So, in general, I need to look at the question first to locate what kind of information I need to process''} (P12). 
They motivated this preference for a task-specific strategy and for locating keywords by the complexity of the process model and the capacity of their memory. They argued that they would forget the details of the process model if reading the model first.
E.g., \textit{``So, I mean, the diagram is relatively complex and a lot of information is out there. If we look at the diagram first, it is possible that we read it from left to right, after we read it, we might even forget what we have read on the left part. So there will be a messy and catastrophes so that is why I choose to identify the key words. Yeah, the verbs are usually very important in phrases. So I believe that we can work out the question successfully.'' }(P6). 
While going through the details of the process model, the participants following a task-specific  strategy mentioned that they considered the relevance between the task, process model, and business rules \textbf{(structure relations-driven}) (e.g., \textit{``I just looked down on the question, see whether there is a relation or not and then I will continue''} (P5). In addition, they revisited the tasks after they read through the whole process to remind them of the tasks. %

We also observed the \textbf{model-driven strategy} in Q1 during focused information foraging, i.e., four participants mentioned they would read through the model first and then read the task.  
They explained that they wanted to get familiar with all the details in the process model and then focus on the question. \textit{``I think is like that like a habit. Or do things like this, it's quite common like you don't familiar with something, you need to get through the total thing, and you need to get familiar with all the diagram, and then you need to focus on the question, what the questions is asking about.''} (P8). All of them mentioned after they read the task, they considered the relevance between the model and the tasks \textbf{(structure relations-driven}).

Despite the differences in strategies (either tasks-specific or model-driven) during the interpretation and analysis of focused foraging information, all experts (15/15) reported that they read through all details when understanding the model and business rules in Q1. E.g. \textit{``I'm going everywhere right now because it's the first time that I'm looking at this. And this is the first question. So I feel like I have to go through everything almost at this point and understand what is happening.''} (P5). 
Meanwhile, while comprehending the model, all experts (15/15) indicated that they would both abstract the key part of the information from the model and remember its logic and structure (\textbf{structure-relations driven strategy}).
E.g., \textit{``I just remember the key parts of the model, so for example, this is not a key part, here, so I know that the key points are these three, so most of the work is done here.'' }(P5). 
The insights align with the findings of eye tracking data, where all the experts had similar efficiency in locating the answer for the local question (Q1)\footnote{The efficiency of locating the answer in Q1:
Text annotation (M = 0.166, SD = 0.126), Diagrammatic (M = 0.116, SD = 0.013) and Linked (M = 0.213, SD = 0.145). }.

\textbf{\textit{Selection and evaluation.}} To select and evaluate the relevant information in different tasks, all experts (15/15) indicated they used different strategies on each task based on their information needs and work habits. These strategies include keyword location, the process of elimination, reverse tracking, and comparison of information and structure between model and task, and within the process model. They expressed that using process elimination, identifying the end of the relevant area, and reverse tracking information can facilitate them downsizing the process to locate the relevant information.
Determining the end of the relevant area instead of reading through the whole model again is the preference in answering local questions.
They also would travel reversely and track the required sequence flows, activities and rules (e.g. \textit{``I will go back to graph reversely to see what kind of path I need to go through to get that result.''} (P12)). 
They explained that these strategies enabled them to better locate relevant information areas and eliminate irrelevant parts, e.g., \textit{``I think I like work with the process of elimination... So that helps you look at smaller number of things, rather than going back again, and again, on the full chart. Every time if I go through every step, it will take a lot more time.''} (P13).   

In Q2 and Q3, we noticed they started to \textbf{optimize their strategies} in foraging and processing information based on different information needs for different integration tasks. All experts (15/15) used a task-specific strategy when working on Q2 and Q3. They expressed that they read questions first and adopted different strategies based on the evaluation of the tasks. E.g., \textit{``it's a different question. It's a different question requires a different strategy.''}(P3). After interpreting and analysing task types, all experts expressed they used a different strategy for local and global questions to select and evaluate the relevant information based on their prior experience in practice. In local questions (Q1 and Q2), the dominant strategy is from end to start (reverse tracking), but in global question Q3, all participants expressed they worked from start to end. E.g., \textit{``actually, yes, question one and two are similar. Question three is a different approach. Question one and two, I work from end to start, but question three I work from start to finish.'' }(P3). 

All of the experts (15/15) expressed that, since they understood the process model in Q1, they directly used the strategy of targeting the keyword to locate the relevant information on the model and rules from their memory in Q2 and Q3. They further stated that they read the model and business rules more efficiently and selectively to evaluate and target the information based on the specific information needs required from the question, instead of foraging for all details. 

Based on expert participants' insights, we assume that the initial focused information foraging during the process model and rule understanding played a vital role in building a mental model in their working memory, which will directly influence their attention when completing the following tasks. 
E.g.,\textit{``for this one, I already knew the process a little bit. And so I could directly search for keywords, determine eligibility for the car, here eligibility, it's quite in the centre of the screen.''} (P15). 
All participants expressed that while locating the relevant area to answer questions, they assessed the information in multiple rounds of comparison and evaluation to ensure their answers were relevant.  
For global question Q3, all participants (15/15) mentioned that evaluating the task made them realise the answer requires more areas and rules when compared to the other two local questions, so they went through the whole process model and rules. 
E.g., \textit{``Because here the question was not concerned with one specific outcome like determine eligibility, but it was concerned with overall the whole process, what is the minimum, so I had to go through the whole process.''} (P15). As task complexity increases, all experts presented similar efficiency in locating the answer in the global question (Q3)\footnote{The efficiency of locating the answer in Q3: 
Text annotation (M = 0.047, SD = 0.035), Diagrammatic (M = 0.046, SD = 0.047), and Linked (M = 0.070, SD = 0.039).}.

\textbf{Task-specific information processing and answering.} 
All experts (15/15) indicated that once they located and confirmed the relevant information in each question, they started to type the answers. 
We observed during this process of answering and assessment, all experts focused on the identified relevant area to synthesize and evaluate relevant information and integrated the information with any prior knowledge or external knowledge. All participants expressed that when they evaluated their answer, they would only target some relevant areas directly based on their memory, but they would not read through all models or rules to confirm again, and they only read it based on the information needed to verify and ensure that they did not overlook anything (e.g., \textit{``for the minimum number, you kind of have to go through the whole process...it's based on the situation so I looked into each rule again quickly, so to make sure that I don't overlook anything and then I will short case.''} (P15).

\section[Conclusion and Outlook]{Conclusion and Outlook}
\label{Sec:discussion}

In this paper, we advanced the understanding of sensemaking practices of process knowledge workers engaged in tasks on integrated modelling of business processes and rule. Utilizing cued-retrospective think-aloud interviews (with eye-tracking as the cue), our study conducted a deep qualitative exploration that revealed diverse sensemaking practices and strategies experts employed during information foraging and information processing. 
Our findings echoed the iterative character noted in prior sensemaking work \cite{Chen2023explor,pirolli2005sensemaking,zhang2020cognitive}, but we uncovered a number of specific observations in expert process workers' behaviours and strategies when performing tasks
based on integrated modeling of business processes and rules, 
namely 
(1) We discovered that sensemaking practices unfold through various processes in information foraging and processing phases, notably highlighting that the information foraging phase consists of two distinct sub-phases, i.e., exploratory and focused information foraging phases.
(2) Our study reveals that expert process workers exhibited a dynamic adaptation to different information needs by optimizing and switching strategies between these phases. Such strategic shifts are crucial in accommodating the varying degrees of complexity and specificity of the information foraged.
(3) Further, we discovered that sensemaking practices are not mutually exclusive but are interrelated and can be concurrently employed. 

While recent advancements in AI-assisted tools offer promising capabilities for process model analysis and decision support \cite{dumas2023ai}, they fall short in capturing the nuanced tacit and experiential knowledge \cite{de2023cognitive,sanzogni2017artificial} that expert knowledge workers apply in real-world scenarios. These limitations can lead to potential misinterpretations of complex business processes and business rules, particularly in contexts where domain-specific expertise is essential but absent. 
Considering these challenges, gaining a deeper understanding of how expert knowledge workers forage for and process information within integrated process models becomes essential. Our work presents such insights, which are crucial for supporting the development of new, more effective process-oriented tools and systems that enhance decision-making processes. Drawing on our findings, we argue that the development of tools and systems in integrated process and rule modeling should be informed by the cognitive mechanisms uncovered through the analysis of sensemaking loops specific to process knowledge workers' expertise. In essence, our research underscores the importance of offering personalized support mechanisms to improve the sensemaking practices of different user groups. 

Our study is not without limitations. 
First, our research focused on integrated process models, yet it did not explore how business rules \cite{figl2018we,vanthienen2004quality} specifically influence expert process workers' information foraging and processing. This limitation presents an opportunity for future research to explore the impact of the full complexity of understanding business rules and impacts on information processes.
Second, we only considered basic constructs in business process models whereas advanced constructs (e.g., events, loops and nesting structures) may introduce further complexities in sensemaking. 
Third, our participant pool of researchers with practitioner backgrounds was limited and the sample size may diminish the generalizability to industrial settings. While the sample size is considered adequate given the exploratory and qualitative nature of our study, which is in alignment  with established qualitative research methodologies that value information richness over large sample sizes \cite{abbad2019exploring,marshall2013does}, a higher number of expert participants may provide better explanatory power.
We consider our combined approach of using cued-retrospective think-aloud interviews with eye tracking as a methodological contribution that can inspire further research. Specifically, 
the methodology outlined in this study can also be applied and extended for research on other conceptual notations and domains with similar related but notationaly distinct artefacts, for example, database constraints and data models.
In our future work, we plan to  
contribute to both the theoretical framework of sensemaking and the practical aspects of designing tools and systems that support effective information foraging, processing and decision-making in multi-artifact information tasks such as business process and rule integration.

We consider our combined approach of using cued-retrospective think-aloud interviews with eye tracking as a methodological contribution that can inspire further research. Specifically, 
the methodology outlined in this study can also be applied and extended for research on other conceptual notations and domains with similar related but notationaly distinct artefacts, for example, database constraints and data models.
In our future work, we plan to  
contribute to both the theoretical framework of sensemaking and the practical aspects of designing tools and systems that support effective information foraging, processing, and decision-making in multi-artefact information tasks such as business process and rule integration.

 \bibliographystyle{splncs04}
 \bibliography{Bibliography}

\begin{thebibliography}{10}
\providecommand{\url}[1]{\texttt{#1}}
\providecommand{\urlprefix}{URL }
\providecommand{\doi}[1]{https://doi.org/#1}

\bibitem{abbad2019exploring}
Abbad~Andaloussi, A., Burattin, A., Slaats, T., Petersen, A.C.M., Hildebrandt, T.T., Weber, B.: Exploring the understandability of a hybrid process design artifact based on dcr graphs. In: EMMSAD 2019, Proceedings 20. pp. 69--84 (2019)

\bibitem{abbad2021exploring}
Abbad~Andaloussi, A., Zerbato, F., Burattin, A., Slaats, T., Hildebrandt, T.T., Weber, B.: Exploring how users engage with hybrid process artifacts based on declarative process models: a behavioral analysis based on eye-tracking and think-aloud. Software and Systems Modeling  \textbf{20},  1437--1464 (2021)

\bibitem{adams1957laboratory}
Adams, J.K.: Laboratory studies of behavior without awareness. Psychological bulletin  \textbf{54}(5), ~383 (1957)

\bibitem{de2023cognitive}
de~Almeida Rodrigues~Gon{\c{c}}alves, J.C., Baiao, F.A., Santoro, F.M., Guizzardi, G.: A cognitive bpm theory for knowledge-intensive processes. Business Process Management Journal  \textbf{29}(2),  465--488 (2023)

\bibitem{bera2011guidelines}
Bera, P., Burton-Jones, A., Wand, Y.: Guidelines for designing visual ontologies to support knowledge identification. Mis Quarterly pp. 883--908 (2011)

\bibitem{bera2019using}
Bera, P., Soffer, P., Parsons, J.: Using eye tracking to expose cognitive processes in understanding conceptual models. MIS quarterly  \textbf{43}(4),  1105--1126 (2019)

\bibitem{boren2000thinking}
Boren, T., Ramey, J.: Thinking aloud: Reconciling theory and practice. IEEE transactions on professional communication  \textbf{43}(3),  261--278 (2000)

\bibitem{chen2016robust}
Chen, F., Zhou, J., Wang, Y., Yu, K., Arshad, S.Z., Khawaji, A., Conway, D.: Robust multimodal cognitive load measurement. Springer (2016)

\bibitem{Chen2023explor}
Chen, T., Demartini, G., Indulska, M., Sadiq, S.: Exploring data workers’ behaviours in data quality discovery. In: ACIS 2023 Proceedings. No.~99 (2023)

\bibitem{chen2020sensemaking}
Chen, T., Sadiq, S., Indulska, M.: Sensemaking in dual artefact tasks – the case of business process models and business rules. In: International Conference on Conceptual Modeling. pp. 105--118 (2020)

\bibitem{chen2018business}
Chen, T., Wang, W., Indulska, M., Sadiq, S.: Business process and rule integration approaches-an empirical analysis. In: BPM Forum 2018. pp. 37--52 (2018)

\bibitem{duchowski2018gaze}
Duchowski, A.T.: Gaze-based interaction: A 30 year retrospective. Computers \& Graphics  \textbf{73},  59--69 (2018)

\bibitem{dumas2023ai}
Dumas, M., Fournier, F., Limonad, L., Marrella, A., Montali, M., Rehse, J.R., Accorsi, R., Calvanese, D., De~Giacomo, G., Fahland, D., et~al.: Ai-augmented business process management systems: a research manifesto. ACM Transactions on Management Information Systems  \textbf{14}(1),  1--19 (2023)

\bibitem{figl2015influence}
Figl, K., Laue, R.: Influence factors for local comprehensibility of process models. International Journal of Human-Computer Studies  \textbf{82},  96--110 (2015)

\bibitem{figl2018we}
Figl, K., Mendling, J., Tokdemir, G., Vanthienen, J.: What we know and what we do not know about dmn. EMISAJ  \textbf{13}, ~2--1 (2018)

\bibitem{gioia2013seeking}
Gioia, D.A., Corley, K.G., Hamilton, A.L.: Seeking qualitative rigor in inductive research: Notes on the gioia methodology. Organizational research methods  \textbf{16}(1),  15--31 (2013)

\bibitem{gruhn2006adopting}
Gruhn, V., Laue, R.: Adopting the cognitive complexity measure for business process models. In: 2006 5th IEEE international conference on cognitive informatics. vol.~1, pp. 236--241. IEEE (2006)

\bibitem{haisjackl2016understanding}
Haisjackl, C., Barba, I., Zugal, S., Soffer, P., Hadar, I., Reichert, M., Pinggera, J., Weber, B.: Understanding declare models: strategies, pitfalls, empirical results. Software \& Systems Modeling  \textbf{15}(2),  325--352 (2016)

\bibitem{haisjackl2018humans}
Haisjackl, C., Soffer, P., Lim, S.Y., Weber, B.: How do humans inspect bpmn models: an exploratory study. Software \& Systems Modeling  \textbf{17},  655--673 (2018)

\bibitem{hart2006nasa}
Hart, S.G.: Nasa-task load index (nasa-tlx); 20 years later. In: Proceedings of the human factors and ergonomics society annual meeting. vol.~50, pp. 904--908 (2006)

\bibitem{knolmayer2000modeling}
Knolmayer, G., Endl, R., Pfahrer, M.: Modeling processes and workflows by business rules. In: Business Process Management, pp. 16--29. Springer (2000)

\bibitem{kurtz2003new}
Kurtz, C.F., Snowden, D.J.: The new dynamics of strategy: Sense-making in a complex and complicated world. IBM systems journal  \textbf{42}(3),  462--483 (2003)

\bibitem{la2013business}
La~Rosa, M., Dumas, M., Uba, R., Dijkman, R.: Business process model merging: An approach to business process consolidation. TOSEM  \textbf{22}(2),  1--42 (2013)

\bibitem{marshall2013does}
Marshall, B., Cardon, P., Poddar, A., Fontenot, R.: Does sample size matter in qualitative research?: A review of qualitative interviews in is research. Journal of computer information systems  \textbf{54}(1),  11--22 (2013)

\bibitem{mendling2008influence}
Mendling, J., Strembeck, M.: Influence factors of understanding business process models. In: International Conference on Busi. Infor. Syst. pp. 142--153 (2008)

\bibitem{mendling2012factors}
Mendling, J., Strembeck, M., Recker, J.: Factors of process model comprehension—findings from a series of experiments. Deci. Supp. Syst.  \textbf{53}(1),  195--206 (2012)

\bibitem{naumer2008sense}
Naumer, C., Fisher, K., Dervin, B.: Sense-making: a methodological perspective. In: Sensemaking Workshop, CHI. vol.~8, pp. 506--513 (2008)

\bibitem{nielsen1994estimating}
Nielsen, J.: Estimating the number of subjects needed for a thinking aloud test. International journal of human-computer studies  \textbf{41}(3),  385--397 (1994)

\bibitem{petrusel2013eye}
Petrusel, R., Mendling, J.: Eye-tracking the factors of process model comprehension tasks. In: 25th International Conference, CAiSE 2013. pp. 224--239. Springer (2013)

\bibitem{petrusel2016task}
Petrusel, R., Mendling, J., Reijers, H.A.: Task-specific visual cues for improving process model understanding. Infor. and Soft. Tech.  \textbf{79},  63--78 (2016)

\bibitem{pirolli2005sensemaking}
Pirolli, P., Card, S.: The sensemaking process and leverage points for analyst technology as identified through cognitive task analysis. In: Proceedings of international conference on intelligence analysis. vol.~5, pp.~2--4 (2005)

\bibitem{rayner1998eye}
Rayner, K.: Eye movements in reading and information processing: 20 years of research. Psychological bulletin  \textbf{124}(3), ~372 (1998)

\bibitem{reijers2011human}
Reijers, H.A., Mendling, J., Dijkman, R.M.: Human and automatic modularizations of process models to enhance their comprehension. Info. Sys.  \textbf{36}(5) (2011)

\bibitem{rolon2008evaluation}
Rol{\'o}n, E., Garc{\'\i}a, F., Ruiz, F., Piattini, M., Visaggio, C.A., Canfora, G.: Evaluation of bpmn models quality-a family of experiments. In: Inter. Confer. on Evaluation of Novel Approaches to Soft. Engineering. vol.~2, pp. 56--63 (2008)

\bibitem{russell1993cost}
Russell, D.M., Stefik, M.J., Pirolli, P., Card, S.K.: The cost structure of sensemaking. In: Proceedings of the INTERACT'93 and CHI'93. pp. 269--276 (1993)

\bibitem{sanzogni2017artificial}
Sanzogni, L., Guzman, G., Busch, P.: Artificial intelligence and knowledge management: questioning the tacit dimension. Prometheus  \textbf{35}(1),  37--56 (2017)

\bibitem{schreiber2024cognitive}
Schreiber, C., Abbad-Andaloussi, A., Weber, B.: On the cognitive and behavioral effects of abstraction and fragmentation in modularized process models. Information Systems  \textbf{125},  102424 (2024)

\bibitem{strauss1998basics}
Strauss, A., Corbin, J.: Basics of qualitative research techniques  (1998)

\bibitem{sweller2011measuring}
Sweller, J., Ayres, P., Kalyuga, S.: Measuring cognitive load. In: Cognitive load theory, pp. 71--85 (2011)

\bibitem{turetken2016effect}
Turetken, O., Rompen, T., Vanderfeesten, I., Dikici, A., van Moll, J.: The effect of modularity representation and presentation medium on the understandability of business process models in bpmn. In: BPM. pp. 289--307 (2016)

\bibitem{van2005uncovering}
Van~Gog, T., Paas, F., Van~Merri{\"e}nboer, J.J., Witte, P.: Uncovering the problem-solving process: Cued retrospective reporting versus concurrent and retrospective reporting. Journal of Experimental Psychology: Applied  \textbf{11}(4), ~237 (2005)

\bibitem{vanthienen2004quality}
Vanthienen, J.: Quality by design: using decision tables in business rules. Business Rules Journal  \textbf{5}(2), ~7 (2004)

\bibitem{vayghan2007internal}
Vayghan, J.A., Garfinkle, S.M., Walenta, C., Healy, D.C., Valentin, Z.: The internal information transformation of ibm. IBM Systems Journal  \textbf{46}(4),  669--683 (2007)

\bibitem{wang2022business}
Wang, W., Chen, T., Indulska, M., Sadiq, S., Weber, B.: Business process and rule integration approaches—an empirical analysis of model understanding. Information Systems  \textbf{104},  101901 (2022)

\bibitem{weick1995sensemaking}
Weick, K.E.: Sensemaking in organizations, vol.~3. Sage (1995)

\bibitem{zhang2020cognitive}
Zhang, P., Soergel, D.: Cognitive mechanisms in sensemaking: A qualitative user study. Journal of the Association for Infor. Sci. and Tech.  \textbf{71}(2),  158--171 (2020)

\bibitem{zhou2023process}
Zhou, H., Zerbato, F., Weber, B., Indulska, M., Sadiq, S.: Do process analysts care about the metadata of event logs? ACIS 2023 Proceedings  \textbf{32} (2023)

\bibitem{zugal2012assessing}
Zugal, S., Pinggera, J., Weber, B., Mendling, J., Reijers, H.A.: Assessing the impact of hierarchy on model understandability--a cognitive perspective. In: Models in Software Engineering. pp. 123--133 (2012)

\bibitem{zugal2015investigating}
Zugal, S., Soffer, P., Haisjackl, C., Pinggera, J., Reichert, M., Weber, B.: Investigating expressiveness and understandability of hierarchy in declarative business process models. Software \& Systems Modeling  \textbf{14},  1081--1103 (2015)

\bibitem{zur2008towards}
Zur~Muehlen, M., Indulska, M., Kittel, K.: Towards integrated modeling of business processes and business rules  (2008)

\end{thebibliography}

 \end{document}